# Nonlinearity-induced photonic topological insulator

Lukas J. Maczewsky[1], Matthias Heinrich[1], Mark Kremer[1], Sergey K. Ivanov[2,3]
Max Ehrhardt[1], Franklin Martinez[1], Yaroslav V. Kartashov[3,4],
Vladimir V. Konotop[5,6], Lluis Torner[4,7], Dieter Bauer[1], and Alexander Szameit[1,*].

[1] Institut für Physik, Universität Rostock, Albert-Einstein-Str. 23, 18059 Rostock, Germany.

[2] Moscow Institute of Physics and Technology, Institutsky lane 9, Dolgoprudny, Moscow Region, 141700, Russia

[3] Institute of Spectroscopy, Russian Academy of Sciences, Fizicheskaya Str., 5, Troitsk, Moscow, 108840, Russia

[4] ICFO-Institut de Ciencies Fotoniques, The Barcelona Institute of Science and Technology, 08860 Castelldefels (Barcelona), Spain

[5] Departamento de Física, Faculdade de Ciências, Universidade de Lisboa, Campo Grande, Lisboa 1749-016, Portugal

[6] Centro de Física Teórica e Computacional, Faculdade de Ciências, Universidade de Lisboa, Campo Grande, Lisboa 1749-016, Portugal

[7] Universitat Politècnica de Catalunya, 08034, Barcelona, Spain

*Correspondence to: alexander.szameit@uni-rostock.de

The hallmark feature of topological insulators renders edge transport virtually impervious to scattering at defects and lattice disorder. In our work, we experimentally demonstrate a topological system, using a photonic platform, in which the very existence of the topological phase is brought about by nonlinearity. Whereas in the linear regime, the lattice structure remains topologically trivial, light beams launched above a certain power threshold drive the system into its transient topological regime, and thereby define a nonlinear unidirectional channel along its edge. Our work studies topological properties of matter in the nonlinear regime, and may pave the way towards compact devices that harness topological features in an on-demand fashion.

**One Sentence Summary:** Nonlinear propagation dynamics induce a topological insulator phase.

With the discovery of topological insulators (TIs) and their experimental realizations[1–6], material science ushered in a new era of physics. In a seemingly paradoxical fashion, solid-state TIs prohibit electrons from traversing their interior, while simultaneously supporting chiral surface currents that are topologically protected from scattering at defects and disorder[7,8]. Soon after the first realizations in condensed matter systems [3,5], topological concepts were implemented across other fields of physics, resulting in the experimental demonstrations of topological dynamics in various different platforms, ranging from microwave systems [9], photonic lattices [10], matter waves [11], acoustics [12], mechanical waves [13], electronic circuits [14] and even polaritons [15]. Among these, optical settings in particular have proven to be suited for realizing topological phenomena [16,17], such as Floquet TI [10,18,19], TI on a silicon platform [20], 4D topological Hall physics [21], Weyl points[22], topological Anderson insulators [23], TIs in synthetic dimensions [24] as well as non-Hermitian topological physics [25,26] and topological quantum physics [27,28].

Currently, researchers are starting to connect topology and nonlinearity[29-31]. An exciting example is the recent demonstration of bulk soliton formation within a topological band gap, a first and impressive step towards nonlinear topological dynamics[29]. Nevertheless, experimental efforts on nonlinear TIs and their corresponding robust edge transport have thus far suffered from one crucial limitation: The respective implementations only allowed for experiments with purely linear edge state dynamics[30,31]. The severity of this constraint, and the need to overcome it, is underpinned by a number of theoretical proposals in condensed matter systems and photonics considering interactions as source of peculiar physics [32]. Nonlinearity may, for example, enable topological Mott insulators [33], interaction-induced topological insulators such as the Kondo insulator [34], non-Abelian topological insulators [35], and even drive the formation of topological solitons [36,37]. Clearly, nonlinearity is the key ingredient for a wide range of novel and fascinating topological physics.

In our work, we theoretically and experimentally demonstrate a nonlinearity-induced photonic TI. We show how nonlinearity can drive an initially topologically trivial system into a transient topological phase, where nonlinearly self-defined chiral edge states exist (see Figure 1). In contrast to soliton formation in arrangements with preexisting topological features[29], it is the action of nonlinearity itself that establishes nontrivial topology here. In our experiments, we employ lattices of coupled optical waveguides as a versatile platform to explore nonlinear physics [38–41]. In particular, we make use of a modified version of an anomalous Floquet topological insulator arrangement [18,19,42], which can exhibit unidirectional edge transport and bulk localization. Interestingly, the Chern number $\mathcal{C}$ of this structure generally remains zero. Instead, the topological phase of this system is adequately described by a winding number $\mathcal{W}$, which counts the number of topologically protected edge states[42].

In the tight-binding regime, the light dynamics in our finite (2+1) dimensional system can be faithfully modelled by the discretized Schrödinger equations[39]

$$i\frac{\partial}{\partial z}a_n(z) + \sum_{\langle m \rangle} H_{m,n}(z)a_m(z) + \gamma|a_n(z)|^2 a_n(z) = 0 \qquad (1)$$

for the field amplitudes $a_n(z)$, where $H_{m,n}(z)$ is the linear tight-binding Hamiltonian, describing the $z$-dependent coupling from site $n$ to a nearest neighbor $m$ (indicated by $\langle m \rangle$ in the sum) in the two-dimensional lattice. In turn, the quantity $\gamma$ describes the strength of our Kerr-type nonlinearity. Evidently, Eq. (1) has a striking similarity to the well-known Gross-Pitaevskii equation, with the temporal evolution being replaced by spatial dynamics in $z$ along the waveguides, and $\gamma > 0$ describing focusing/positive nonlinearity [43,44]. We consider a (2 + 1) dimensional spatially discrete system, which does not exhibit chiral edge states, and will remain topologically trivial for low-intensity beams. However, at sufficient excitation powers, the term $\gamma|a_n(z)|^2 a_n(z)$ may become significant in Eq. (1). As we will show in our experiments, this can in fact drive the system into a topologically non-trivial phase and promote the formation of nonlinear chiral edge states.

We start our experiments by demonstrating the underlying principle using two interacting waveguides forming a directional coupler[45] (see Fig. 2a). Whereas such a coupler comprised of identical waveguides exhibits the characteristic sinusoidal intensity oscillation and the associated periodic full power transfer between the waveguides, introducing a detuning between their effective refractive indices forces a certain fraction of light to remain in the initially excited guide at all times. In other words, in the linear regime, the propagating light can no longer transfer fully to the second waveguide, and the maximum reached in the second waveguide remains well below the total input power. However, when launching high-power light into the guide with lower refractive index, a positive nonlinearity ($\gamma > 0$) may establish an intermittent phase-matching with the neighboring guide. As a result, a substantially larger fraction of power can therefore be transferred from the excited to the other waveguide (see Fig. 2b). By this means, the light transfer ratio can be tuned under the influence of nonlinearity. The photonic lattices employed in our experiments were fabricated by means of femtosecond laser direct inscription[46], indeed show this behavior when excited with intense laser pulses. The observed power dependence of the output intensity is shown in Fig. 2b: For low input powers ($P_{\text{lin}} = 100\text{kW}$ peak power), only about 44% of the light is transferred out of the excited channel. We were able to seamlessly tune this transfer efficiency up to a value of 73% for $P_{\text{NL}} = 2.9\text{MW}$. In Figs. 2c and 2d, the simulated transfer dynamics in a two waveguide system as a function of the input power illustrates how such an increased transfer ratio can be achieved despite the focusing nature of the nonlinearity by judiciously tailoring the length of the interaction region. The nonlinear switching presented in this experiment is the basis for our implementation of a nonlinear photonic topological insulator.

Our topological structure is comprised of optical waveguides arranged in a bipartite square lattice (see Fig. 3a; the two site species are indicated by light and dark circles, respectively). The individual channels are selectively brought into evanescent contact with one another so as to form a cyclic discrete hopping pattern. At any distinct propagation step $j$, light from each waveguide is partially transferred with ratio $t$ to only one specific nearest neighbor. Additionally, the on-site potential (i.e. the refractive index) of the waveguides is modulated for each successive step. The entire four-step scheme is illustrated in Fig. 3b, where the waveguide diameter symbolizes the strength of detuning and its change along the direction of propagation. This Floquet driving protocol serves to establish a lattice that is periodic in the transverse ($x, y$) and the longitudinal ($z$) direction, with driving period $Z$. Notably, the system enters a non-trivial topological phase when the transfer ratio in each step exceeds $t > 0.5$ [18]. Above this critical transfer ratio corresponding to a topological phase transition, the Chern number $\mathcal{C}$ remains zero, while the winding number assumes the value of $\mathcal{W} = 1$ (Ref. [42]).

While the transfer ratio of each driving step determines the topological phase of the system, the behavior of the band structure depends on the micro-dynamics of the electric field. In this vein, we analyze the linear dynamics of all four individual steps $j$ of length $Z/4$, by considering the $(k_x, k_y)$-space of the time-discretized linear bulk Hamiltonian. By applying a Bloch ansatz to the infinitely extended real space Hamiltonian of Eq. (1) of the bipartite lattice displayed in Fig. 3a, we find

$$H_\text{B}(\boldsymbol{k}, z) = \sum_{j=1}^{4} \begin{pmatrix} \delta^{(1)}(z) & c_j(z)e^{i\boldsymbol{b}_j \boldsymbol{k}} \\ c_j(z)e^{-i\boldsymbol{b}_j \boldsymbol{k}} & \delta^{(2)}(z) \end{pmatrix}, \qquad (2)$$

where the vectors $\{\boldsymbol{b}_j\}$ are defined as $\boldsymbol{b}_1 = -\boldsymbol{b}_3 = (d/2, 0)$ and $\boldsymbol{b}_2 = -\boldsymbol{b}_4 = (0, d/2)$, with the transverse lattice constant $d$ (see Fig. 3c). During the $j^\text{th}$ step of the period the coupling $c_j = c$, while the other three values are set to zero. The detuning $\delta^{(1)}$ of the on-site potential is set to value $\delta$ in step 1 and 4 and remains zero for step 2 and 3. Its counterpart $\delta^{(2)}$ is fixed to $\delta$ in step 2 and 3 and set to zero for step 1 and 4 (as visualized in Fig. 3b,c). We start our analysis by choosing the coupling coefficient as $c = 1.5\frac{\pi}{Z}$ and a detuning value of $= 3.2\frac{\pi}{Z}$, which correspond to the parameters used for the detuned waveguide pair from Fig. 2.[47]. In this case, during each step, the transfer ratio $t$ is only 44%, and as a consequence the system remains well within the topologically trivial phase, as both Chern number and winding number vanish ($\mathcal{C} = \mathcal{W} = 0$). The corresponding quasi-energy spectrum $\varepsilon(\boldsymbol{k})$ (Ref. [42]) of the system is shown in Fig. 3d, where only dispersive bands and a clearly visible band gap exists. Particularly, the edge quasi-energy spectrum $\varepsilon(k_x)$ of the truncated array shows no trace of chiral edge states.

The situation changes dramatically, when nonlinearity comes into play. The dynamics between linear coupling $c$ and nonlinear self-focusing can be described by an effective nonlinear detuning value $\delta_\text{NL}^\text{eff} = \delta_\text{lin} + \delta_\text{NL}(|a_n|^2)$[43,48], where $\delta_\text{lin}$ is the permanent detuning of the sites introduced during the fabrication process ($\delta^{(1)}$ and $\delta^{(2)}$). In the case of the preliminary experiment described above, $\delta_\text{NL}^\text{eff} = 1.2\frac{\pi}{Z}$. Crucially, the virtually instantaneous nature of the Kerr electronic response guarantees that any such propagation effects are strictly local in our system. The Floquet theorem is, in general, not applicable to nonlinear differential equations such as (1) because the dynamics, which feeds back into the nonlinear part of the Hamiltonian, does not necessarily have the same time-periodicity as the linear part of the Hamiltonian [49]. However, by carefully tuning the distances over which two evanescent waveguides couple, we ensure that the nonlinear dynamics remains periodic in the evolution (i.e., $z$ in our case), and so does the nonlinear Hamiltonian. Following the approach of [50], it is therefore possible to calculate a meaningful quasi-energy spectrum of the Floquet lattice with $\delta_\text{NL}^\text{eff}$, the result of which is shown in Fig. 3e. The nonlinear effective detuning results in an increased transfer ratio $t > 0.5$, which leads to a change into a local non-trivial topological phase. In this vein, the band retains its essential characteristics, a chiral edge state emerges (marked blue), indicating that the system has indeed been driven into a topological phase by the nonlinear light evolution. Accordingly, for this effective nonlinear coupling, the local winding number can be computed to be $\mathcal{W} = 1$ [18,19,42]. Importantly, this effective spectrum is only present as long as the power is sufficient to drive the required effective detuning. As such, the resulting chiral edge states exist only by the grace of nonlinear light dynamics.

Combining all these ingredients, we set out to observe the actual formation of nonlinear chiral edge states in an experimental setting. To this end, we fabricated extended samples comprised of 36 waveguides, with the four partial coupling steps of one and two full Floquet period fabricated in samples up to $Z = 15$cm length. The required detuning between the two constituent sites of each unit cell was implemented by an appropriate choice of writing speeds. In each coupling step, the interaction region was tailored such that an identical effective linear coupling was achieved. When launching low-power light at $P_{\text{lin}} = 100$kW (i.e. the transfer ratio at each coupling step is $t < 0.5$), into an edge waveguide, light diffracts into the bulk of the lattice and especially in all surrounding edge sites, indicating its topologically trivial phase (see Fig. 4a,c). In contrast, as the input peak power is increased to $P_{\text{NL}} = 3.5$MW (i.e. $t > 0.5$), this bulk diffraction becomes suppressed, and the desired chiral edge state forms (Fig. 4b,d). Similar to its counterparts in conventional linear TI, this state is topologically protected against scattering. The observed edge state occupation for two driving periods, determined as ratio $I_{\text{edge}}/I_{\text{total}}$ of the intensity of the four involved edge lattice sites (indicated by the dashed blue outline in Fig. 4d) over the total intensity in the lattice, clearly demonstrates the transition to the topological phase as the power is gradually increased (Fig. 4e). In contrast, a bulk excitation (Fig. 4d-j) contracts towards its initially excited site at higher launched powers (Fig. 4g,i), resulting in a sudden increase of the bulk state occupation, calculated as the ratio $I_{\text{bulk}}/I_{\text{total}}$ of the intensity at the linear bulk state lattice sites (indicated by dashed blue outline in Fig. 4i) over the total intensity. Crucially, this contraction does not arise merely from a nonlinear cancellation of coupling. Instead, it is a signature of suppressed bulk transport in conventional TIs. The loop-like trajectory of the nonlinear excitation in the system at hand is indicated by a yellow arrow in Fig. 4g,i. Despite the transient nature of the underlying topological phase, the edge light transport is robust, and can persist even in the presence of imposed coupling disorder and artificial defects – as long as the power of the propagating beam is sufficient to induce the required amount of effective coupling. In other words, nonlinearly induced topologically protected edge states transport can be terminated at will by introducing losses so as to reduce the intensity below the phase transition threshold. Alternatively, this reverse transition can also be brought about by the overall decrease of peak intensity due to propagation losses or dispersive pulse broadening. Notably, the ability to terminate topological protection at will, and at a location encoded by the initially injected power or pulse pre-chirp, addresses one of the main challenges hindering the application of topological systems, namely how to efficiently extract signals at the end of their topologically protected journey.

In our work, we proposed and experimentally demonstrated a genuine nonlinearity-induced photonic topological insulator. We showed that this structure, despite its topologically trivial behavior under linear conditions, high-power beams experience self-induced topological edge protection as well as a suppression of bulk transport. Based on an integrated-optical platform, these findings open a new experimental avenue towards developing a more holistic understanding of topological insulators. Indeed, many fascinating questions come to mind: What exactly will be the impact of nonlinearity in systems exhibiting fermionic time-reversal symmetry [51]? How can gain, loss, and non-Hermiticity in general, be efficiently harnessed to tailor nonlinear wave packet evolution [52]? Does the presence of nonlinearity in non-trivial topologies in fact enhance or postpone disorder-induced localization mechanisms [53]? The tools to experimentally tackle these topics are now within reach, and will drive the exploration of the scientific as well as technological aspects of nonlinear topology in all kinds of wave mechanical systems, whether in the photonic, acoustic, optomechanical, polaritonic, atomic, or even entirely new domains.

**Acknowledgments:** The authors would like to thank C. Otto for preparing the high-quality fused silica samples used for the inscription of all photonic structures employed in this work.

**Funding:** The authors acknowledge funding from the Deutsche Forschungsgemeinschaft (grants SCHE 612/6-1, SZ 276/12-1, BL 574/13-1, SZ 276/15-1, SZ 276/20-1) and the Alfried Krupp von Bohlen and Halbach foundation. V.V.K. acknowledges supported from the Portuguese Foundation for Science and Technology (FCT) under Contract no. UIDB/00618/2020. Y.V.K. and S.K.I. acknowledge funding of this study by RFBR and DFG according to the research project no. 18-502-12080.

**Author contributions:** All authors contributed significantly, discussed the results and co-wrote the manuscript.

**Competing interests:** The authors declare no competing interests.

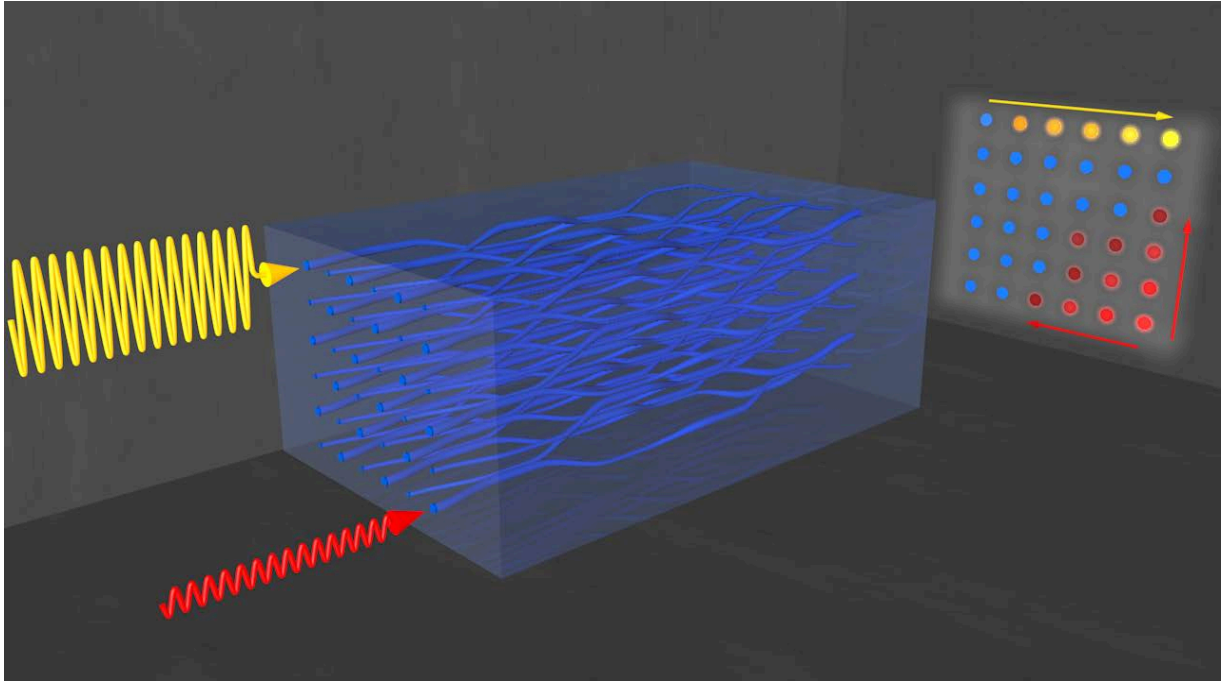

**Fig. 1:** Nonlinearity-induced photonic topological insulator. Low-power edge excitations (red) undergo substantial diffraction into the bulk of the blue waveguide structure, indicating a topologically trivial regime. In contrast, injecting high-power light (yellow) gives rise to a self-guided unidirectional edge state travelling along the perimeter of the structure. In this topologically nontrivial phase, leakage into the bulk is suppressed and the signal becomes resilient towards scattering.

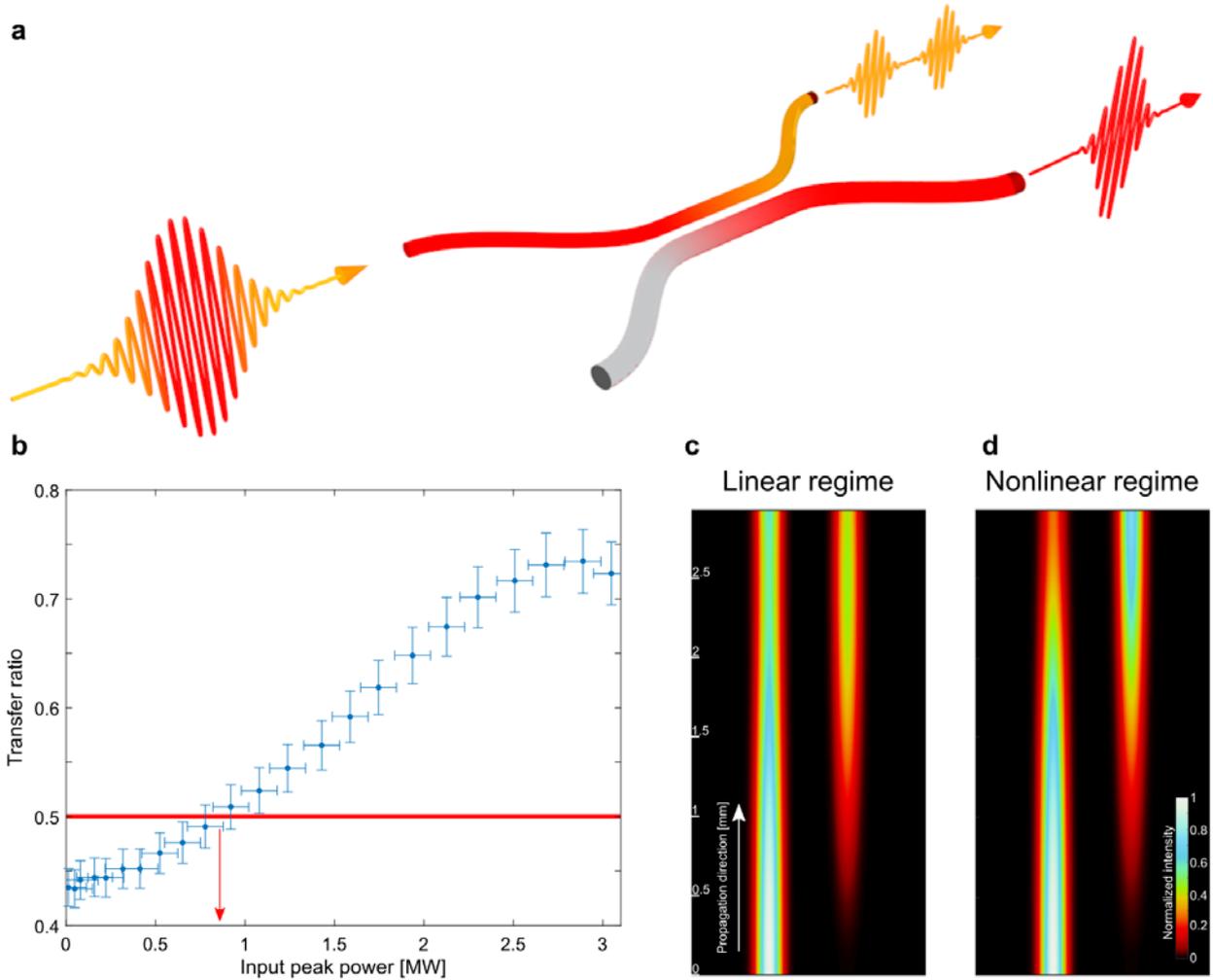

**Fig. 2:** Nonlinear directional detuned coupler. (a) Schematic: An intense laser pulse is launched into the lower-refractive-index channel of a detuned coupler. A focusing Kerr nonlinearity allows the high-intensity core (red) of the pulse to momentarily compensates the detuning, whereby light is transferred to the high-refractive-index waveguide. In contrast, the lower-intensity slopes of the pulse (yellow) exhibit linear coupling behavior. (b) Measured power-dependent transfer efficiency in a detuned coupler. The horizontal red line indicates the threshold value required to establish the topologically non-trivial phase in our modified anomalous Floquet topological insulator. The red arrow represents the corresponding input peak power for this transition. (c,d) Simulated light propagation in a detuned coupled waveguide structure for (c) low and (d) high intensities.

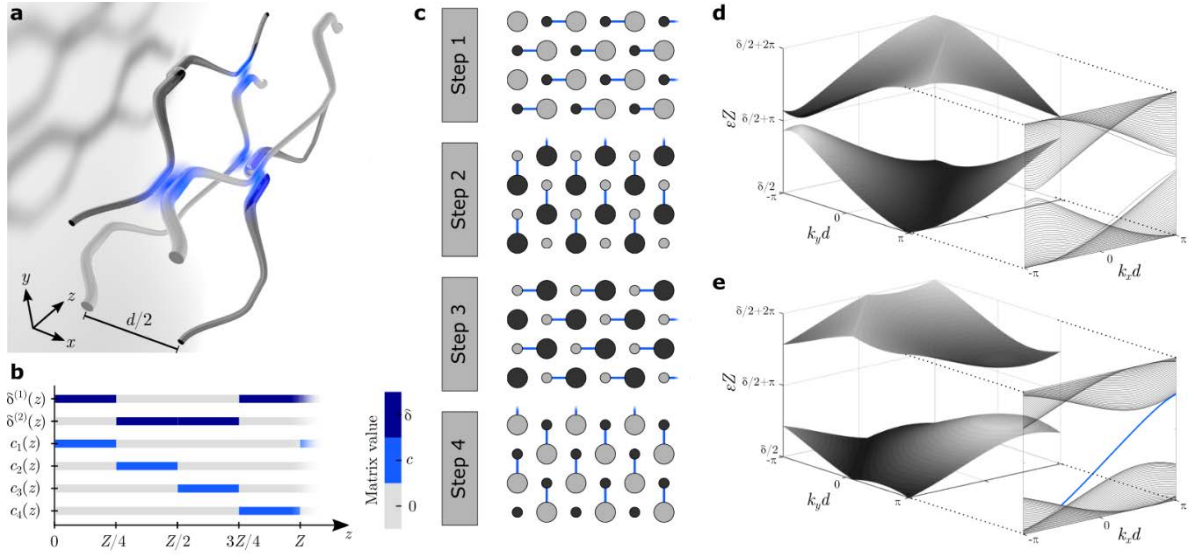

**Fig. 3:** Driving Protocol. (a) Schematic of the waveguide-based implementation of the coupling sequence detailed in (b,c). (b) Visualization of the changing parameters of Equation (2), which characterize the driving protocol. (c) The two different waveguides species (light/dark grey) are allowed to interact within the blue-shaded regions. In addition, the on-site potential of each waveguide is modulated along the propagation direction, as indicated by different waveguide diameters. (d,e) The three-dimensional bulk band structure of the photonic structure is periodic in the quasi-momenta $k_x$, $k_y$ as well as in the quasi-energy $\varepsilon$, and exhibits a pronounced band gap. Depending on the effective detuning, the two-dimensional edge band structure features (d) only a band gap in the trivial regime if no nonlinearity is present, and (e) a chiral edge state (blue solid line) connecting the bands in the topological phase, when nonlinearity compensates the detuning of the onsite potential.

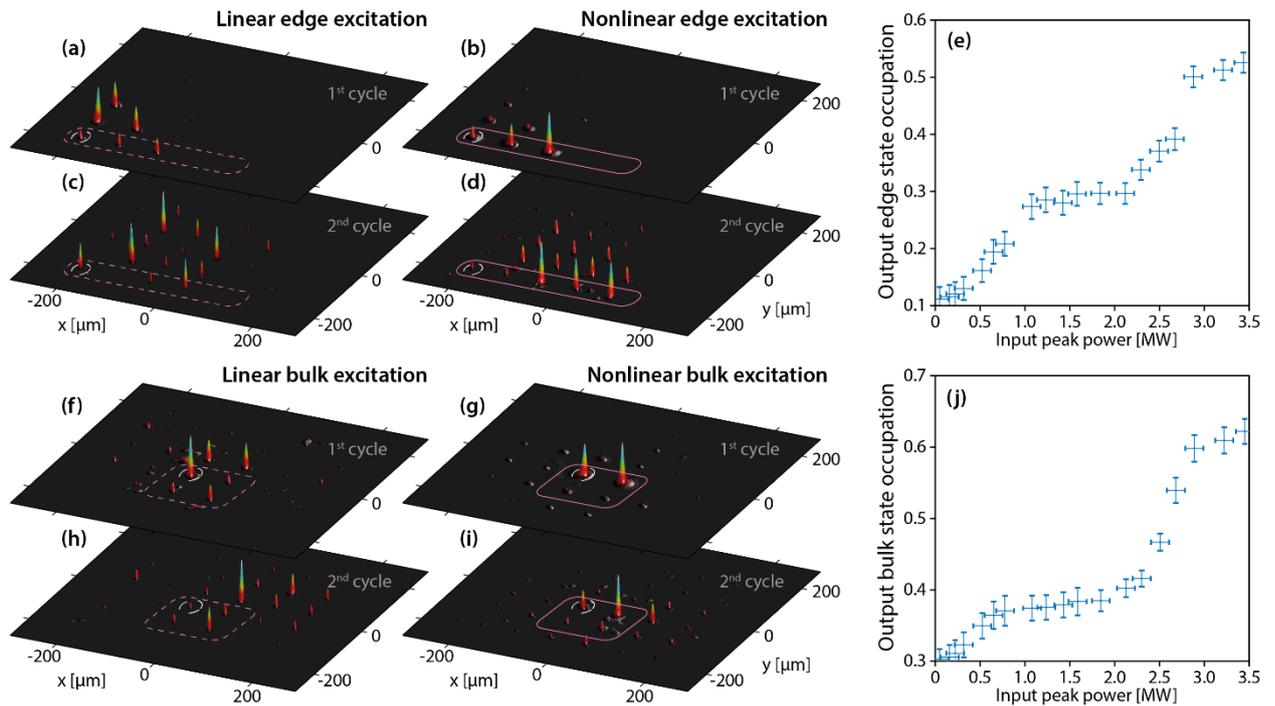

**Fig. 4:** Experimental observation of nonlinearly induced topological dynamics. (a-d) Output intensity distribution for single-site excitation at orange marked lattice edge site for (a,b) one driving period and (c,d) two driving periods. (a,c) Bulk-diffractive behavior of a low-power edge excitation. (b,d) Unidirectional edge transport at high powers of 3.1MW, indicated by orange arrows. (e) Edge occupation as function of the injected power after two driving periods. Above the nonlinearly induced phase transition, an edge state clearly emerges. (f-i) Output intensity distribution for single-site excitation at orange marked bulk lattice site for (f,g) one driving period and (h,i) two driving periods. (f,h) Bulk dynamics around the excitation site for low-power excitation. (g,i) Suppression of bulk diffraction above the nonlinear phase transition, confirmed by (j) a marked increase of the bulk state occupation.